\documentclass[]{raa}            

\usepackage{graphicx,times}             
\usepackage{natbib}
\usepackage{amssymb,amsmath}
\usepackage{threeparttable,array}
\bibpunct{(}{)}{;}{a}{}{,}

\usepackage[UTF8, scheme=plain]{ctex} 

\usepackage[pagebackref=true]{hyperref}

\begin{document}

\title{
  Pulsar Backend for 21 CentiMeter Array: Implementation of Data Acquisition and Initial Results
}

\volnopage{Vol.0 (20xx) No.0, 000--000}      
\setcounter{page}{1}          

\author{
  Yukai Zhou (周宇凯) \inst{1}
  \and
  Junhua Gu (顾俊骅) \inst{2,3,*} \footnotetext{$*$ Corresponding Authors, these authors contributed equally to this work.}
  \and
  Mengyao Xue (薛梦瑶) \inst{3}
  \and
  Faxin Shen (沈发新) \inst{4,5}  
  \and
  Jian Li (李健) \inst{5,6,7}
  \and
  Qiuyang Fu (付秋阳) \inst{3,5}  
  \and
  Cijie Zhang (张赐杰) \inst{3}  
  \and
  Youling Yue (岳友岭) \inst{3,*}
  \and
  Weiwei Zhu (朱炜玮) \inst{2,8}
  \and
  Kejia Lee (李柯伽) \inst{1,2,9}
  \and
  Renxin Xu (徐仁新) \inst{1,9}
}

\institute{
Department of Astronomy, School of Physics, Peking University, Beijing 100871, China
\and
State Key Laboratory of Radio Astronomy and Technology, National Astronomical Observatories, Chinese Academy of Sciences, A20 Datun Road, Chaoyang District, Beijing, 100101, China; {\it jhgu@nao.cas.cn}
\and
National Astronomical Observatories, Chinese Academy of Sciences, Beijing 100012, China; {\it ylyue@nao.cas.cn}
\and
Yunnan Observatories, Chinese Academy of Sciences, 650216 Kunming, China
\and
School of Astronomy and Space Science, University of Chinese Academy of Sciences, Beijing 100049, China
\and
Xinjiang Astronomical Observatory, Chinese Academy of Sciences, Urumqi 830011, China;
\and
Xinjiang Key Laboratory of Microwave Technology, Urumqi 830011, China
\and
Institute for Frontiers in Astronomy and Astrophysics, Beijing Normal University, Beijing 102206, China
\and
Kavli Institute for Astronomy and Astrophysics, Peking University, Beijing 100871, China
\vs\no
{\small Received 20xx month day; accepted 20xx month day}
}

\abstract{
  We implemented a data acquisition system for 21 CentiMeter Array (21CMA),
  enabling baseband observations 
  targeting pulsars and fast radio bursts.  
  Based on the Radio Frequency System-on-Chip (RFSoC) platform,
  the new backend is capable of
  instantaneously covering the effective bandwidth from 50 to 350 MHz,
  with multi-board synchronization achieved at the timescale of the sampling clock.
  We observed PSR B0329$+$54 with a single station to verify the signal path integrity;
  then solved phase relations of multiple station pairs
  using bright persistent radio sources like Cas A and Cyg A;
  using these phase solutions, a multiple-station coherently beamformed observation
  of PSR B0329$+$54 was carried out, showing a signal-to-noise ratio of 699.09 for a 2.5-hour observation with eight stations,
  opening up a possibility of tied-array low-frequency pulsar observations on 21CMA.
  \keywords{
    instrumentation: interferometers
    ---
    (stars:) pulsars: general
    ---
    (stars:) pulsars: individual PSR B0329$+$54
  }
}

\authorrunning{Y. Zhou, et al.}            
\titlerunning{Pulsar Backend for 21CMA: Implementation of Data Acquisition and Initial Results}  

\maketitle


%
%
\section{Introduction}           
\label{sect:intro}
  Pulsars, discovered in the 1960s \citep{Bell_1968}, are widely considered
  highly magnetized, rapidly rotating compact stars. 
  The number of known pulsars has reached about 4000 \citep{psrcat}.
  Pulsars radiate electromagnetic pulses across a wide spectrum covering radio to gamma rays \citep{Thompson_2004}.
  The radio radiation shows a steep spectrum, i.e., in the radio band the pulsar is brighter at lower frequencies, with the spectral indices typically about
  $ -1.6 $ \citep{Jankowski_2017}. 
  Observing pulsars at lower frequencies (e.g. 50 -- 350 MHz) provides opportunities to search for pulsars with 
  steep spectra (e.g., \cite{J1552+5437_Pleunis_2017}).

  Low-frequency pulsar observations also help probe properties of the interstellar medium.
  The interstellar medium interacts with the pulsar radio wave,
  causing the dispersion \citep{handbook_of_pulsar_astronomy}.
  Observing at lower frequency helps  
  constrain the dispersion measurement (DM) precisely, which provides observational data for modeling the Milky Way's electron 
  density distribution \citep{NE2001_1, NE2001_2, YMW16, NE2025_Ocker_2026} and reduces the DM noise of pulsar timing \citep{DM_PTA_KJLee_2014}. 
  Additionally, low-frequency pulsar observations offer approaches for probing
  the ionosphere \citep{ionospheric_scintillation_Fallows_2014, ionospheric_scintillation_Fallows_2020}
  and the interplanetary medium \citep{interplanetary_scintillation_Fallows_2023},
  and studying scintillation \citep{scintillation_arc_Stinebring_2001, interstellar_scintillation_PSR_B0355+54_Xu_2018, fast_radio_burst_scintillation_arc_Wu2023}.

  Observations of pulsars at lower frequencies are usually carried out with antenna arrays and
  require dedicated data recording backends with higher time resolution.
  Examples of existing instruments include 
  Murchison Widefield Array \citep{MWA_Tingay_2013} voltage capturing system \citep{MWAVCS_Tremblay_2015},
  Low Frequency Array \citep{LOFAR_van_Haarlem_2013},
  and
  Canadian Hydrogen Intensity Mapping Experiment \citep{CHIME_Amiri_2022} pulsar backend \citep{CHIME_Pulsar_Amiri_2021}.
  With these pulsar backends,
  wide-field low frequency radio interferometers are capable of covering the observable sky in a short time,
  e.g.,
  MWA's Southern-Sky MWA Rapid Two-Metre (SMART) pulsar survey \citep{SMART_1, SMART_2} 
  covers the entire sky south of declination +30° using 80-minute observations and fewer than 100 hours in total;
  LOFAR Tied-Array All-Sky Survey (LOTAAS) \citep{LOTAAS_overview} covers the sky above declination 0° using a total of 1953 individual pointings.
  Although these surveys consume less observation time compared to similar surveys carried out with large-diameter single dishes or arrays,
  they typically require extensive computing resources to form a large number of coherent beams to maximize their sensitivity.

  The 21 CentiMeter Array (21CMA) \citep{21CMA_Zheng_2016, 21CMA_Huang_2016}
  is a wide-field low-frequency radio array 
  and now a pathfinder for the Square Kilometre Array (SKA). In this paper, we present 
  a new 21CMA data acquisition system 
  designed for 50 -- 350 MHz frequency coverage and $\mu$s-level time resolution,
  enabling pulsar surveys and timing observations.
  This paper is organized as follows: 
  \autoref{section:basic_information_of_21CMA} describes properties of 21CMA, 
  \autoref{sect:system_design} describes the design of the new data acquisition system,
  \autoref{sect:observation_result} presents  
  laboratory tests, initial results of
  a single station observation of PSR B0329$+$54,
  signal delay calibration of stations, and
  a coherently beamformed observation of PSR B0329$+$54,
  \autoref{sect:discussion} provides a brief discussion,
  and the paper concludes in \autoref{sect:conclusion}.

\section{Basic Information of 21CMA}
\label{section:basic_information_of_21CMA}

\begin{figure}[!htbp]
  \centering
  \includegraphics[width=0.5\textwidth]{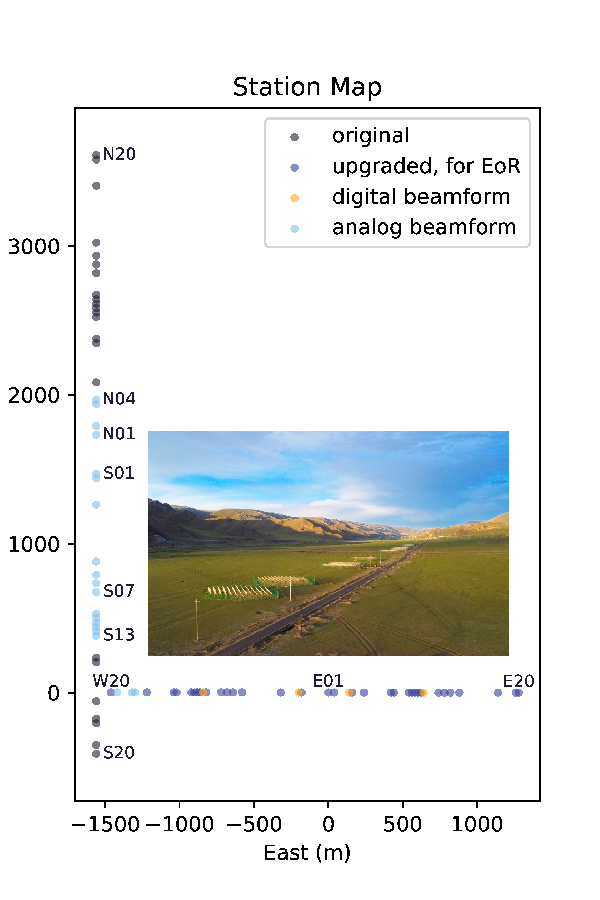}
  \caption{{
      The geometric positions of 21CMA stations. 
      X- and Y-axis are along the east-west and south-north direction, respectively. 
      Positions are relative to the E01 station.
      Each station is denoted with filled dots and the color indicates the current status of each station. 
      The station E21 is beyond the scale and not shown. 
      The photo is a bird's-eye view along the 21CMA east-west arm.
  }}
  \label{fig:21cma_station_map}
\end{figure}

  The 21CMA was initially designed to detect the
  power spectrum of the Epoch of Reionization (EoR). It has been collecting data
  from the north celestial pole region since 2008.
  It is composed of 81 hexagon-shaped stations,
  a photograph and
  the coordinates of the stations are shown in \autoref{fig:21cma_station_map}.
  Each station has 127 log-periodic dipole array (LPDA) antennas and each antenna receives a single linear polarization in the horizontal direction. 
  The electric signal picked from the feed point of the antenna is amplified by a per-antenna low-noise amplifier. 
  The analog beamformer is used to form the station beam, where the signal of each antenna is physically delayed using cables and switches. 
  The delay time is tuned for each antenna such that the combined signal forms a beam pointing towards the target. 
  After signals of all antennas are combined using a radio frequency (RF) power combiner, 
  the combined signal is amplified and transmitted to the central processing room using radio-over-fiber (RFoF) technology.
  
Since 2020, 21CMA has been undergoing upgrades to:
  (1) shift its field of view away from the north celestial pole and cover a larger area of the sky;
  (2) improve time resolution to observe pulsars and fast radio bursts (FRBs).

  The tunable in-station beamforming is implemented 
  to steer the single station beam and extend the field of view.
  Two approaches are being tested:
  (i) sample each antenna using analog-to-digital converters (ADCs) and perform FPGA-based digital beamforming \citep{21CMA_in_station_digital_beamform_Gu_2026},
  which has been deployed on four stations of E03, E13, W02, W09;
  (ii) attach an extra remote-controlled analog delay and attenuation tuner to each antenna and
  perform analog beamforming based on power combiners,
  which is being deployed on stations S01 -- S13, N01 -- N07 and
  will be described by Lee et al. (in prep.). 
  The newly installed beamforming system is capable of instantaneously covering the band of up to 50 -- 350 MHz.

  Given the bandwidth, observing transient signals further requires $\mu$s-level time resolution,
  hence it requires a new data acquisition system and a new data processing pipeline.
  This paper describes the design and implementation of the data acquisition system.

\section{Design of the data acquisition system}
\label{sect:system_design}

\begin{table}[!htbp]
  \centering
  \caption{{
      Specifications of this Data Acquisition System \label{table:specification}
  }}
  \begin{threeparttable}
  \begin{tabular}{ccc}
    \hline\hline
                            & Typical capability$\dagger$  & 21CMA specification                                                         \\
    \hline
    Sample rate             & 1 -- 5 Gsps                    & 1.6 Gsps with 2 $ \times $ decimation                                                       \\
    Analog input bandwidth  & 0 -- 6 GHz                     & 0 -- 400 MHz                                                                                \\
    Instantaneous bandwidth & /                              & \begin{tabular}[c]{@{}c@{}}0 -- 400 MHz (output),\\50 MHz -- 350 MHz (effective)\end{tabular} \\
    Resolution              & 14 bits                        & 8 bits                                                                                      \\
    Analog input count      & 8 (per chip)                   & 20 (recent), 81 (long-term)                                                                 \\
    Digital interface       & 100 GbE$^{*}$ and/or PCIe$^{**}$            & 100 GbE                                                                                     \\
    \hline\hline
  \end{tabular}
  \begin{tablenotes}
            \item [$\dagger$] Typical capabilities are of RFSoC ZU47DR devices.
            \item [*] Gigabit Ethernet
            \item [**] Peripheral Component Interconnect Express
        \end{tablenotes}
      \end{threeparttable}
\end{table}

RFSoC\footnote{\url{https://www.amd.com/en/products/adaptive-socs-and-fpgas/soc/zynq-ultrascale-plus-rfsoc.html}},
a radio frequency platform introduced by Xilinx,
is chosen as the basis of this data acquisition system.
RFSoC consists of ARM cores (processing system, PS),
FPGA (Field Programmable Gate Array; also called programmable logic, PL),
radio frequency data converters (RFDC)
and other components.
The RFDC is a hardware intellectual property (IP) core integrated into the RFSoC family of FPGAs, 
providing both analog-to-digital and digital-to-analog conversion capabilities. 

The hardware specification of the RFSoC platform adopted by us is summarized in \autoref{table:specification},
where the configuration we used for 21CMA is also specified. We designed the firmware to control
the ADC and Ethernet data streaming, and the control interface is implemented
using a customized Linux operating system running on PS.  The implementation of ADC interface and
synchronization with the external 1 pulse per second (1PPS) signal is described in 
\autoref{subsection:adc}. The designs of high-speed Ethernet interface and
PS-based control interface are presented in \autoref{subsection:data_processing} and \autoref{subsection:ps_image}, respectively.
The related source code is publicly available\footnote{\url{https://github.com/fxzjshm/rfsoc_data_acquisition}, branch ``microphase-t510-21cma''}.

\subsection{ADC Setup and Multi-board Synchronization}
\label{subsection:adc}

Due to the low bandwidth requirement of low-frequency radio observations,
direct sampling is used to reduce the complexity of this data acquisition system.
For the 21CMA north-south arm, currently the raw sample rate is set to 1.6 Gsps
and downsampling by a factor of 2 is performed in RFDC,
which gives an effective sample rate of 800 Msps.

\begin{figure}[!htbp]
  \centering
  \includegraphics[width=1\textwidth]{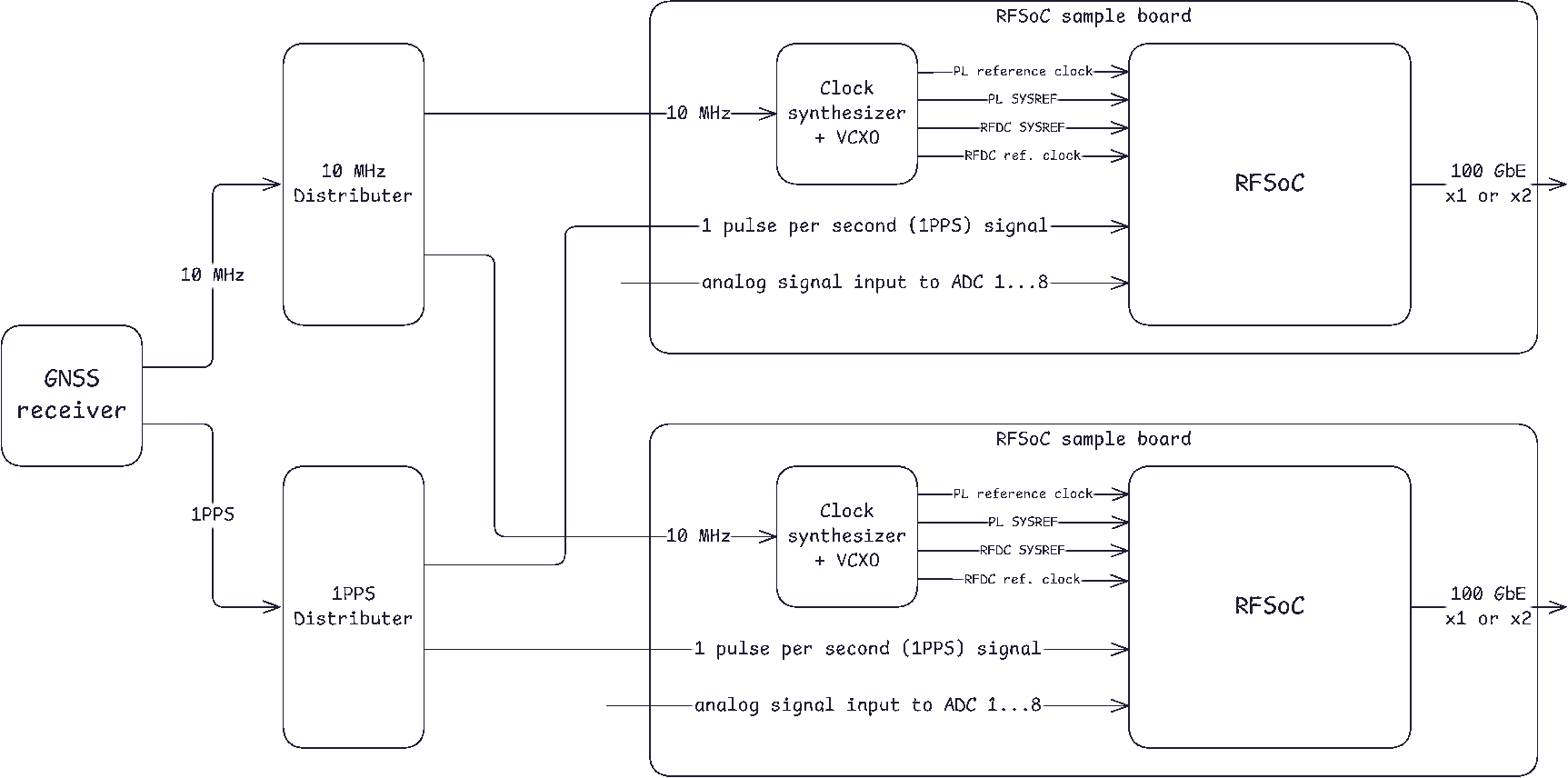}
  \caption{{
    Clock tree of 21CMA RFSoC-based data acquisition system.
  }}
  \label{fig:rfsoc_system_design}
\end{figure}

\begin{figure}[!htbp]
  \centering
  \includegraphics[width=1\textwidth]{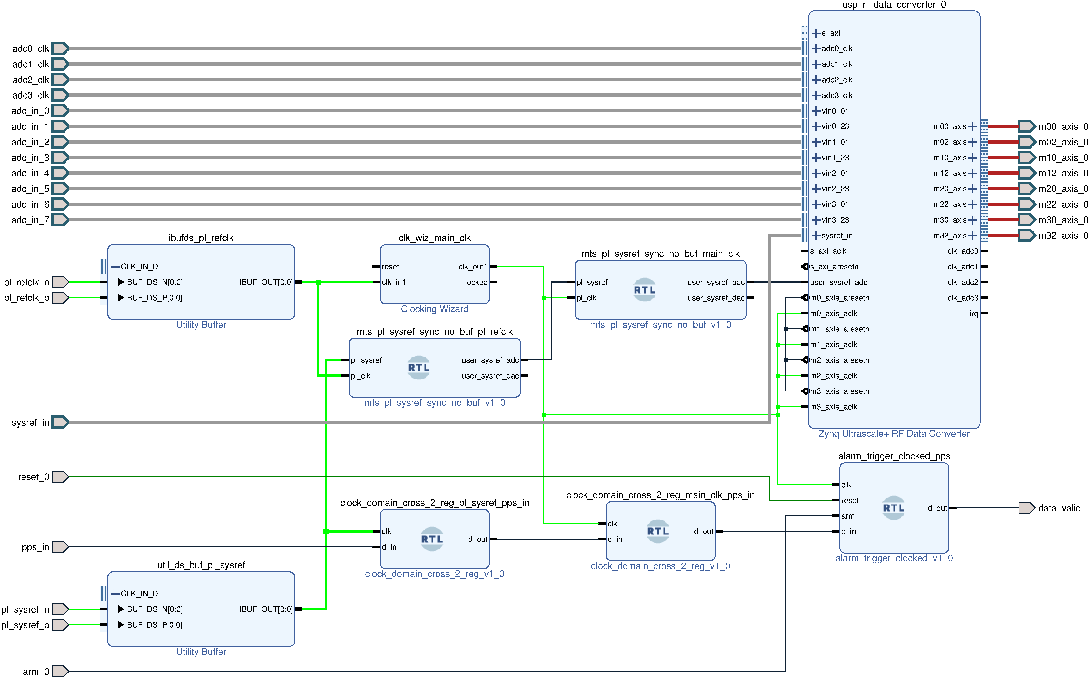}
  \caption{{
    Block design of ADC-related connections in FPGA firmware.
    Some signals are omitted for brevity.
  }}
  \label{fig:fpga_firmware_adc_part}
\end{figure}

Synchronization among multiple ADC tiles on a single chip 
is achieved through the multi-tile synchronization (MTS) procedure provided by Xilinx;
synchronization across multiple acquisition boards 
is accomplished by aligning them to a global 10 MHz reference clock and a global 1 pulse per second (1PPS) signal.
\autoref{fig:rfsoc_system_design} shows the clock tree between RFSoC boards, and
\autoref{fig:fpga_firmware_adc_part} shows the connection inside FPGA firmware.
At 21CMA, in the absence of a hydrogen clock,
a global navigation satellite system (GNSS) receiver is used instead
to provide the 10 MHz reference clock and the 1 pulse per second signal;
these two signals are distributed to all sample boards
using distribution amplifiers.

The on-chip data transfer clock, the ADC sampling clock, 
and the SYSREF signal used for synchronization
are all provided by the clock synthesis chip. 
When generating these secondary clocks, 
the clock synthesis chip relies on phase-locked loops (PLLs), 
which involves multiplying and dividing the input signal. 
Conventional dividers generally suffer from output phase ambiguity, 
making the configuration of the frequency synthesizer chip non-trivial. 
A so-called ``nested zero-delay mode'' is typically employed to 
eliminate the phase ambiguity of the divider output signals.

\begin{figure}[!htbp]
  \centering
  \includegraphics[width=0.8\textwidth]{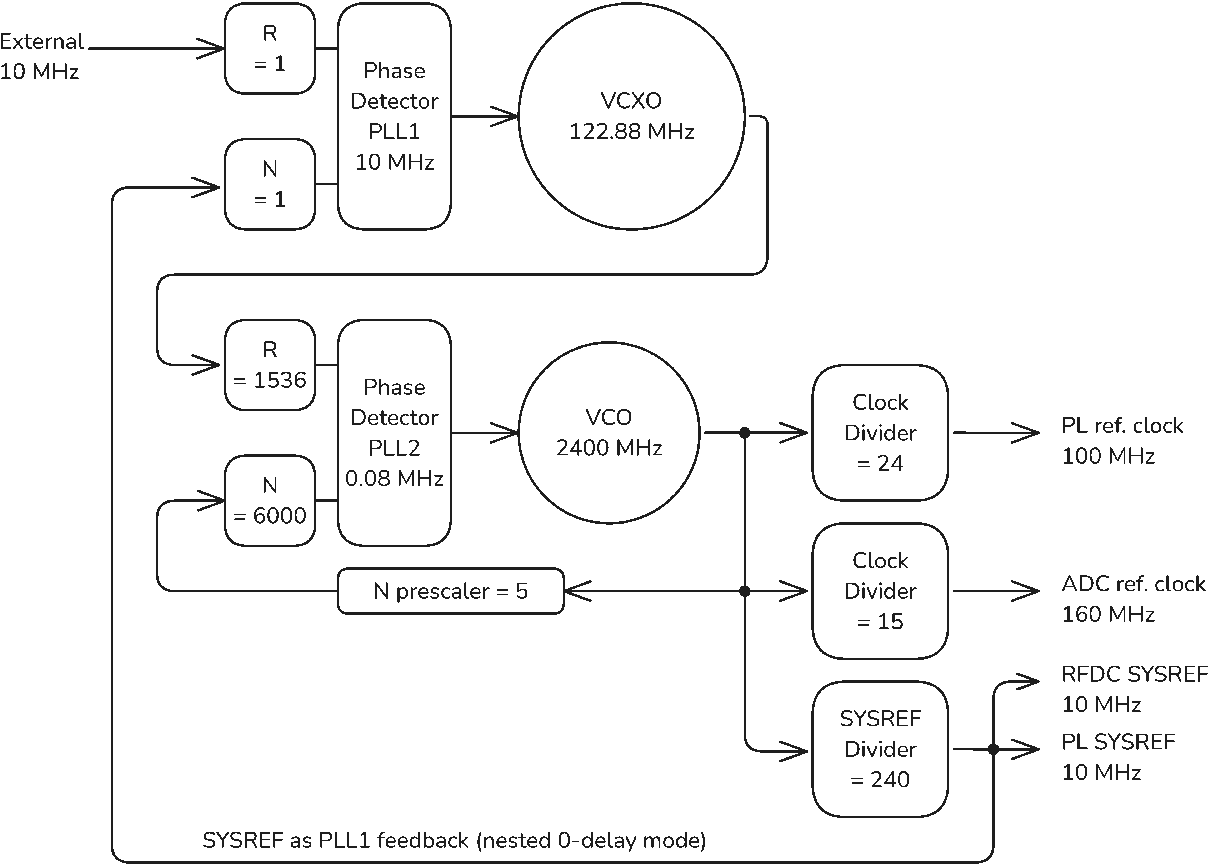}
  \caption{{
    Example configuration inside LMK04828 clock synthesizer for MicroPhase ANTSDR T510 board.
  }}
  \label{fig:T510_LMK04828_config}
\end{figure}

\autoref{fig:T510_LMK04828_config} shows an example configuration of the clock synthesizer on MicroPhase ANTSDR T510 board,
which has an LMK04828 chip as the clock synthesizer chip and 122.88 MHz VCXO (voltage-controlled crystal oscillator).
There are two crucial factors for resolving divider phase ambiguity:
(1) use the lowest-frequency secondary clock as feedback for
the first-stage PLL (PLL1);
(2) ensure the greatest common divisor of the frequencies of all output clocks and the input reference clock is equal to the frequency of the input reference clock.
By comparing SYSREF (10 MHz) phase with the 10 MHz reference clock and matching their frequencies, 
the output SYSREF is aligned with the input without clock divider phase ambiguity;
all other secondary clocks, being integer multiples of 10 MHz, are thus automatically aligned and unambiguous.
Once the secondary clocks of all acquisition boards are mutually phase-aligned 
and the MTS procedure is completed, 
sample clocks of ADCs on all boards become aligned.

To achieve multi-tile and multi-board synchronization,
the following actions are performed:
Analog SYSREF is utilized by the RFDC core to eliminate the phase uncertainty of internal clock dividers.
PL SYSREF is captured by PL reference clock and then by the AXI-Stream output clock as required by RFSoC documentation,
shown in \autoref{fig:fpga_firmware_adc_part};
this helps measure and compensate for the asynchronous first-in-first-out (FIFO) queue between RFDC and PL.
One pulse per second (1PPS) signal is used to align the beginning of the data streams of each sample board,
i.e., data streams are considered valid only after the next rising edge of 1PPS signal
after receiving a ``start working'' notification from PS.
Previously, 1PPS signal was directly captured by RFDC AXI-Stream output clock, typical rate of which is 100 MHz;
however, data misalignment of one AXI-Stream clock was noticed
during the stress tests in laboratory;
this is possibly caused by the rise time of the 1PPS signals,
and to relax this requirement, now 1PPS is captured by 10 MHz SYSREF and then by the AXI-Stream output clock,
as shown in \autoref{fig:fpga_firmware_adc_part}.

With nested 0-delay mode aligning ADC sample clocks,
and the SYSREF signal and the 1PPS signal aligning the beginning of data streams,
the physical time of samples in all data streams can be properly aligned;
furthermore, with the deterministic latency feature in RFDC driver,
the beginning of each data stream is relatively static with respect to the rising edges of 1PPS signal.
Verifications of these features are shown in \autoref{section:lab_test} and \autoref{section:verification_of_synchronization_and_initial_calibration}.

\subsection{Data Processing}
\label{subsection:data_processing}

\subsubsection{Channelization}

Considering multiple requirements of pulsar and transient observations show below,
it is decided that frequency channelization of 21CMA pulsar backend is performed on GPUs,
while the FPGAs serve as a digitizer and a data provider with a UDP interface.

There are three possible implementations of signal channelization:
(a) channelization with FPGA,
(b) channelization with server (using CPU and/or accelerator cards like GPU), and
(c) two-stage scheme where FPGA performs a coarse channelization and server further refines the channel.

For this pulsar backend,
channelization on FPGA cannot meet the channel count requirement.
Analog fiber delays between stations are up to about 4 kilometers or even larger,
which corresponds to approximately $ 1.3 \times 10^4 $ samples when sample boards are operating at 800 Msps.
To achieve acceptable signal coherence after channelization,
an FFT of size 65536 points or larger is required.
Even if coarse delay compensation is done in FPGA to solve signal delay,
which already increases complexity,
avoiding dispersion measure (DM) smearing of pulse signals 
also requires numerous channels.
Implementing channelization with a large number of channels in FPGA
significantly increases complexity of FPGA firmware,
as it may exhaust resources on the chip if implemented naively, 
or may require fine-tuning of fixed-point arithmetic and pipelining;
neither approach is affordable. 

Two-stage channelization is not considered either, because
it requires oversampled polyphase filterbanks to smooth edges of coarse channels,
which increases complexity in FPGA and pushes the digital data rate even higher.

We decided that channelization is done on GPU and skipped on FPGA,
considering that GPU performance is increasing more rapidly than FPGA performance,
programming on GPUs is much easier than programming on FPGAs, and
65536-point FFT is acceptable on GPUs.
As a result, each Ethernet frame is designed to carry the time-domain voltage data from a single ADC.

\subsubsection{Ethernet output}

UDP is chosen to enable Ethernet multicast, 
allowing the acquired data stream to be reused by multiple processing pipelines
and thereby reserving the interfaces for supporting 
simultaneous observation backends targeting different scientific objectives.

UDP packet payload format is chosen to be compatible with
the ROACH2\footnote{\url{https://github.com/casper-astro/casper-hardware/tree/master/FPGA_Hosts/ROACH2}}-based backend\footnote{\url{https://github.com/zhuyangh/fastmb_roach2}}
currently used for Five-hundred-meter Aperture Spherical Telescope (FAST);
this format includes an unsigned 64-bit integer as packet counter and 4096 signed 8-bit time-domain samples from the same ADC;
Ethernet packets like this exceed the common maximum transfer unit (MTU) limit of 1500 bytes
and are called jumbo frames. Jumbo frames reduce the overall count of Ethernet packets
and reduce the overhead of processing packet headers in network interface cards and operating system network stack.

100G Ethernet UDP block\footnote{\url{https://github.com/casper-astro/mlib_devel/tree/m2021a/jasper_library/hdl_sources/onehundred_gbe}}
from CASPER project \citep{Hickish_CASPER_2016} is utilized for transmitting such packets;
it provides a wrapper around the Xilinx 100G CMAC IP core and
implements UDP encapsulation for high-throughput data transmission.

Then the main complexity in this part lies in the strategy for 
outputting ADC data streams to the network interface.
On RFSoC, hardware limitations require 
sending 8 or more data streams through only one or two 100 GbE interfaces. 
All 8 ADCs inside one FPGA output raw data simultaneously in each data transfer clock cycle;
however, at any moment only one or two UDP frames are being transmitted, 
which requires implementing a scheduling mechanism in the FPGA to coordinate the data transfer.

Here we present two strategies of data stream management,
one using a double buffer strategy based on dual port random access memory (RAM) and
the other using an AXI-Stream switch.

\subsubsection{Data Management Strategy 1: Double Buffer Based on Dual Port RAM}

\begin{figure}[!htbp]
  \centering
  \includegraphics[width=1\textwidth]{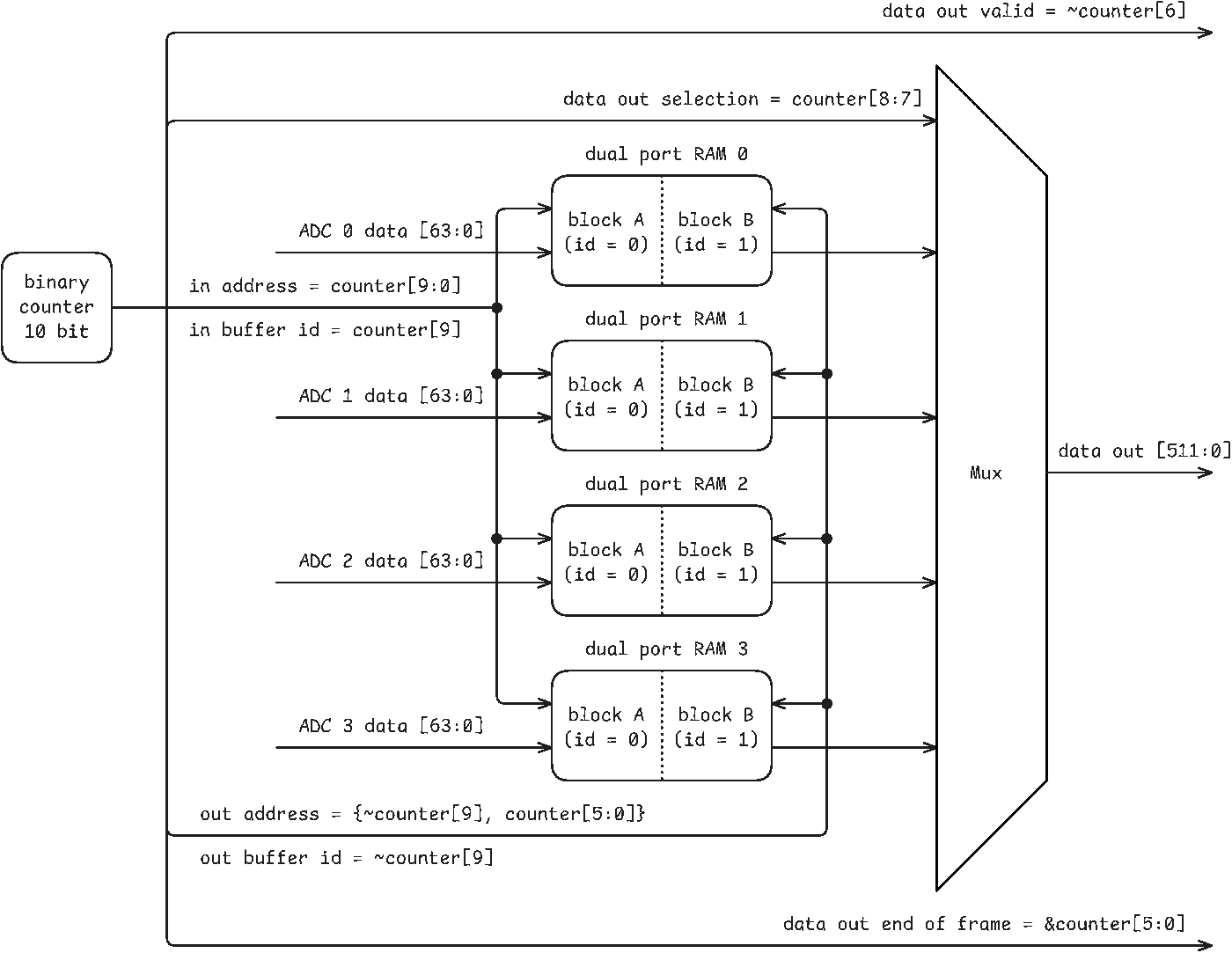}
  \caption{{
    Processing procedures for sending four data streams using one 100 GbE interface
    with double buffer strategy based on dual port RAM.
  }}
  \label{fig:design_2_data_stream}
\end{figure}

\begin{figure}[!htbp]
  \centering
  \includegraphics[width=1\textwidth]{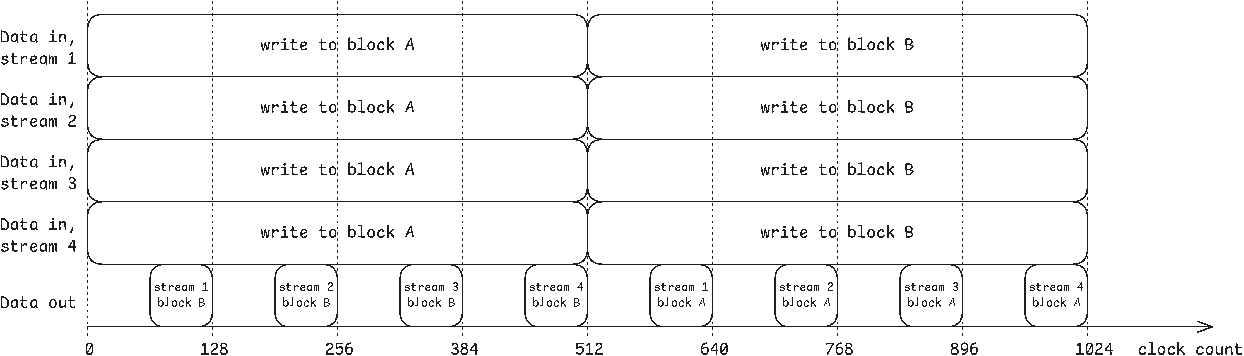}
  \caption{{
    Timeline of processing procedures for sending four data streams using one 100 GbE network interface
    when using double buffer strategy based on dual port RAM.
      X-axis is count of clock.
      The period of these procedures is 1024 clocks
      and exactly one period from start to end is shown in this figure.
  }}
  \label{fig:design_2_data_stream_time}
\end{figure}

\autoref{fig:design_2_data_stream} shows the double buffer strategy based on dual port RAM.
This structure is derived from the firmware of
the pulsar backend for 4 $ \times $ 4.5 m L-band antenna at Guizhou Normal University (Huang et al. in prep.)
developed using CASPER toolflow for ZCU111 and TQ47DR boards \citep{CASPER_TQ47DR_LiJian_2025},
implemented in 2024;
a later work \cite{WangZhao_RFSoC_2025} also uses this strategy.

To support UDP payload format used in this system,
each RAM is configured to have a size of 8192 bytes and is split into two buffers of size 4096 bytes, later referenced as A and B;
when input data is being written to one buffer, output data is read from the other one,
controlled by signals shown in \autoref{fig:design_2_data_stream} as ``in buffer id'' and ``out buffer id''.

When writing to one buffer is complete, the writing process switches to the other buffer;
since the input stream is 64 bits wide and output stream is 512 bits wide, both controlled by the same clock,
this design can iterate over data streams and transmit them,
up to 8 data streams if no extra overhead is introduced by other procedures.
Timeline of input and output events is shown in \autoref{fig:design_2_data_stream_time}.

A drawback of this double buffer strategy, or generally the CASPER toolflow, is that
all logic components use the same clock;
considering that the input is 64 bits wide and output is 512 bits wide,
and extra clocks are required to insert a header containing packet counter before samples,
it is not possible to send eight data streams through one network interface
using this double buffer strategy.
This strategy was sufficient previously since it was designed for ZCU111 and TQ47DR,
on which two 100 GbE interfaces are available through an FMC expansion card or on board,
so eight data streams are sent out using two interfaces;
however, on sample boards with only one 100 GbE interface like MicroPhase ANTSDR T510,
this is no longer appropriate.

\subsubsection{Data Management Strategy 2: AXI-Stream Switch}

\begin{figure}[!htbp]
  \centering
  \includegraphics[width=1\textwidth]{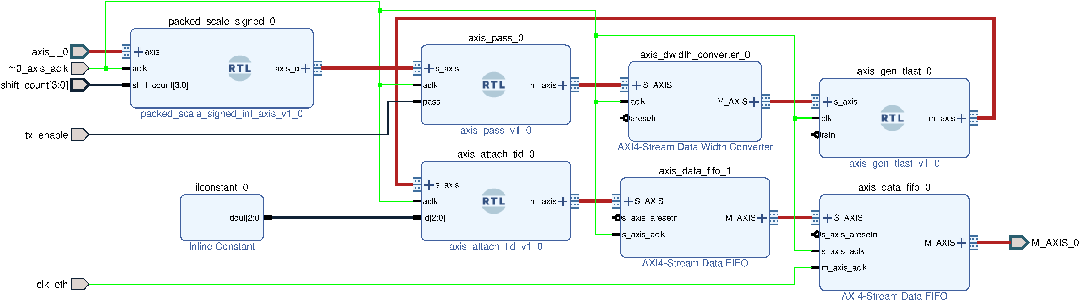}
  \caption{{
    Block design of
    the processing procedure of a single data stream
    using the AXI-Stream switch strategy.
    Some signals are omitted for brevity.
  }}
  \label{fig:design_3_data_process_i}
\end{figure}

\begin{figure}[!htbp]
  \centering
  \includegraphics[width=1\textwidth]{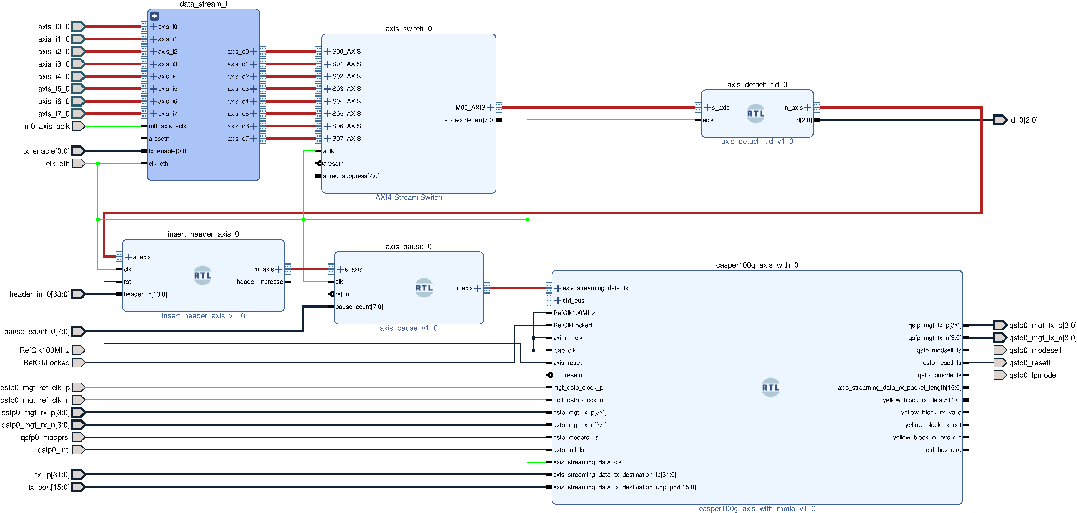}
  \caption{{
    Block design of
    processing procedures for sending eight data streams using one 100 GbE network interface
    when using strategy based on AXI-Stream switch.
    Some signals are omitted for brevity.
  }}
  \label{fig:design_3_data_process}
\end{figure}

To support more data streams in one Ethernet interface,
this firmware also implements a strategy based on AXI-Stream switch.
\autoref{fig:design_3_data_process_i} shows the processing procedure of one data stream.
\autoref{fig:design_3_data_process} shows the overall connections.

  \autoref{fig:design_3_data_process_i} shows the processing procedures for each data stream.
  Samples from the ADC are packed into 8 $ \times $ 16-bit signals.
  A \verb|packed_scale_signed_int_axis| module scales these samples into 8-bit resolution as required by the UDP payload format used.
  \verb|axis_pass| module controls whether a data stream passes through this module or is ignored based on the control signal from the 1PPS module.
  The AXI4-Stream Data Width Converter is a Xilinx IP and is set to
  convert the data width from 8 $ \times $ 8 bits to 64 $ \times $ 8 bits
  to match the data width of the 100 GbE block.
  \verb|axis_gen_tlast| generates an AXI-Stream ``tlast'' signal
  every 64 successful data transfers to mark the end of an AXI-Stream frame
  and also the end of the payload of a UDP packet.
  \verb|axis_attach_tid| module attaches a stream ID to the AXI-Stream data stream using ``tid'' signal.
  Two AXI4-Stream Data FIFOs are UltraRAM-based and block-RAM-based respectively;
  the former is for data caching and
  the latter is for clock domain crossing
  from ADC AXI-Stream clock to the Ethernet transmission clock.
  UltraRAM is used to solve the congestion of block RAM;
  the block-RAM-based FIFO should be large enough to contain a whole UDP buffer,
  otherwise AXI-Stream switch will waste clock cycles waiting for ``tlast'' signal and even cause data corruption.

  \autoref{fig:design_3_data_process} shows the collection of all data streams.
  Eight data streams are collected by an AXI-Stream switch.
  This switch is set to ``packet mode'' so that
  it arbitrates and switches to another input source only at the end of a packet and
  ensures packet completeness.
  \verb|axis_detach_tid| module detaches stream ID to select the destination IP, destination port and
  related header information from the PS-PL interconnection registers.
  A packet counter is generated and inserted before the sample data by the \verb|insert_header_axis| module.
  Finally, a complete UDP payload is sent to the CASPER 100 GbE block.
  \verb|axis_pause| module pauses data transfer for several clocks after each packet
  to mitigate possible buffer overflow inside the CASPER 100 GbE block.

\subsection{Processing System Image}
\label{subsection:ps_image}

To maximize adaptability, embedded Linux is chosen instead of bare-metal applications
to control the FPGA part.
A hybrid method is used to build the operating system of the processing system:
for the kernel, PetaLinux is used to simplify building a custom Linux kernel instead of building Linux from scratch;
for userspace, Ubuntu for MPSoC\footnote{\url{https://ubuntu.com/download/amd}}
is used to form a standard Linux environment,
which simplifies program dependency management
since package managers like APT and pip are available.

The PS-PL AXI interconnect is utilized to control registers inside the FPGA,
e.g., the destination IP and UDP port of eight data streams, and the ``start capturing'' command;
a custom register map is provided to
vhdmmio\footnote{\url{https://github.com/abs-tudelft/vhdmmio/}},
and vhdmmio generates a memory-mapped input/output (MMIO) block inside FPGA
that configures these registers within the FPGA.
For RFDC registers, the official RFDC driver\footnote{\url{https://github.com/Xilinx/embeddedsw/tree/master/XilinxProcessorIPLib/drivers/rfdc/}}
is utilized to achieve multi-tile synchronization (MTS) and multi-board synchronization;
minor modifications to RFDC driver are made
to disable DAC synchronization and enable the deterministic latency feature.

PS is also responsible for writing bitstreams to the FPGA and
writing configurations to registers in the clock synthesizer chip using protocols like SPI.
This enables
flexibility
for runtime-switchable sample modes;
if in the presence of an RF switch matrix,
this can also enable the reuse of sample boards for multiple ultra-wideband receivers
of some large radio telescopes.

\section{Test and Deployment}
\label{sect:observation_result}

\subsection{Synchronization Test in Laboratory}
\label{section:lab_test}

In the laboratory testing stage, two MicroPhase ANTSDR T510 boards are connected as shown in \autoref{fig:rfsoc_system_design},
except that the source of 1PPS and 10 MHz is replaced by a rubidium frequency standard. 
The ADC0 of each board is connected to a separate 1PPS output of the same rubidium frequency standard 
with a 20 dB attenuator and a power splitter.

\begin{figure}[!htbp]
  \centering
  \includegraphics[width=0.5\textwidth]{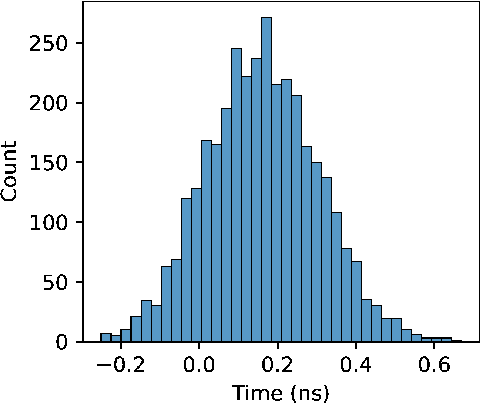}
  \caption{{
    Histogram of the time delays measured in the multi-board synchronization test in the laboratory.
    X-axis: delay time in nanosecond. Y-axis: count of time delays in range.
  }}
  \label{fig:rfsoc_sync_test_lab_3462}
\end{figure}

By measuring the timing positions of the 1PPS signal captured by two boards across several restarts,
both the deterministic latency relative to 1PPS and the multi-board synchronization can be checked.
No inconsistency has been found among 3462 tests performed.
Two series of digitized voltage data are correlated in the frequency domain
and the slope of correlation phase against frequency is fitted using the RANSAC method \citep{RANSAC}
to derive delay time.
The histogram of measured delay times is shown in \autoref{fig:rfsoc_sync_test_lab_3462}.
The standard deviation of delay time is 0.14 ns, about an order of magnitude lower than the sampling period,
indicating that multi-board synchronization is achieved.
The mean delay is 0.16 ns,
representing a constant time delay that is likely caused by the power splitters or the cables used.


\subsection{Pulsar observation using a single station}

\begin{figure}[!htbp]
  \centering
  \includegraphics[width=0.5\textwidth]{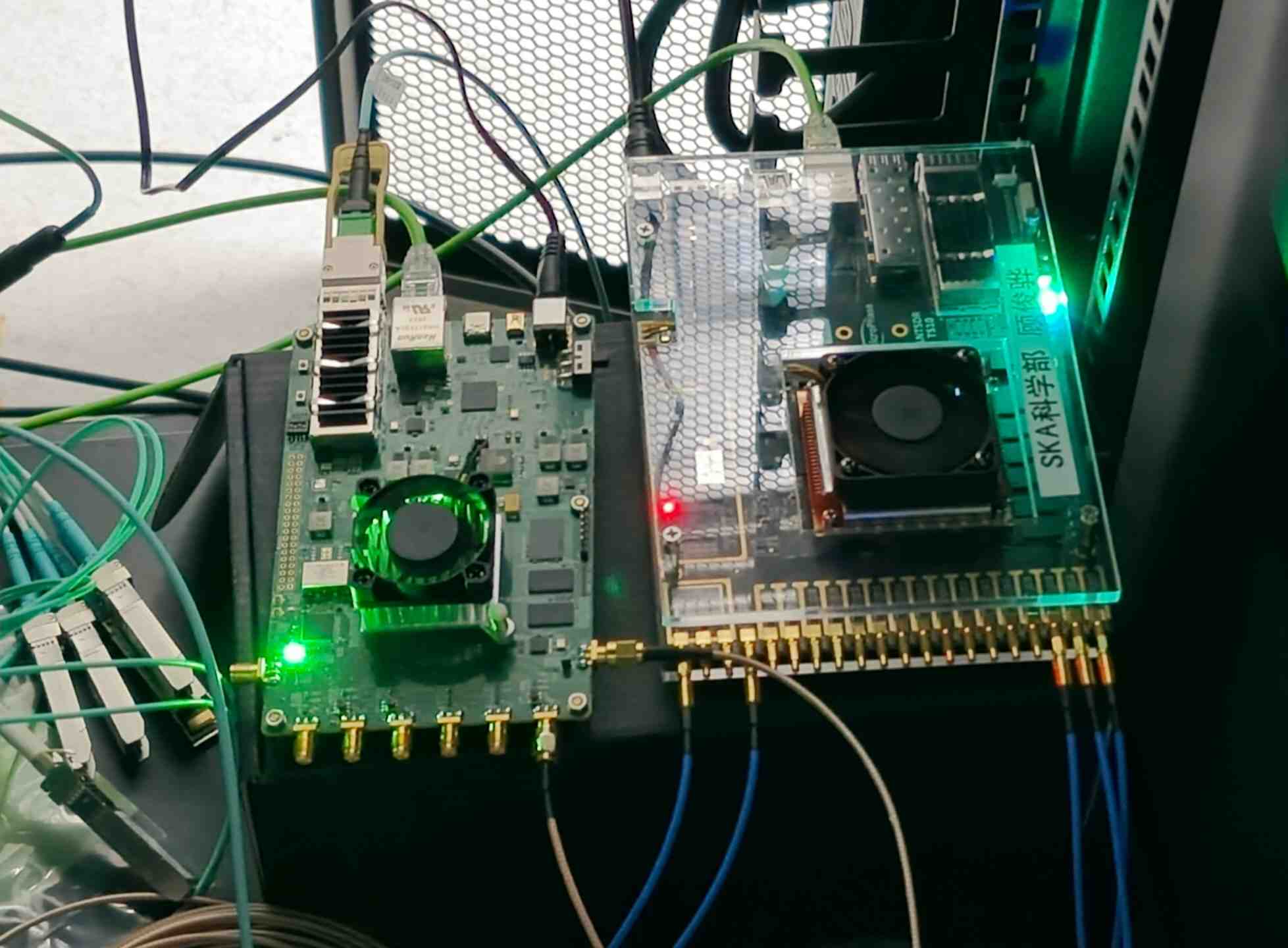}
  \caption{{
    Some of RFSoC sample boards deployed at 21CMA.
    Left: GRALIC (4T4R).
    Right: MicroPhase ANTSDR T510.
  }}
  \label{fig:deployed_rfsoc}
\end{figure}

\begin{figure}[!htbp]
  \centering
  \includegraphics[width=1\textwidth]{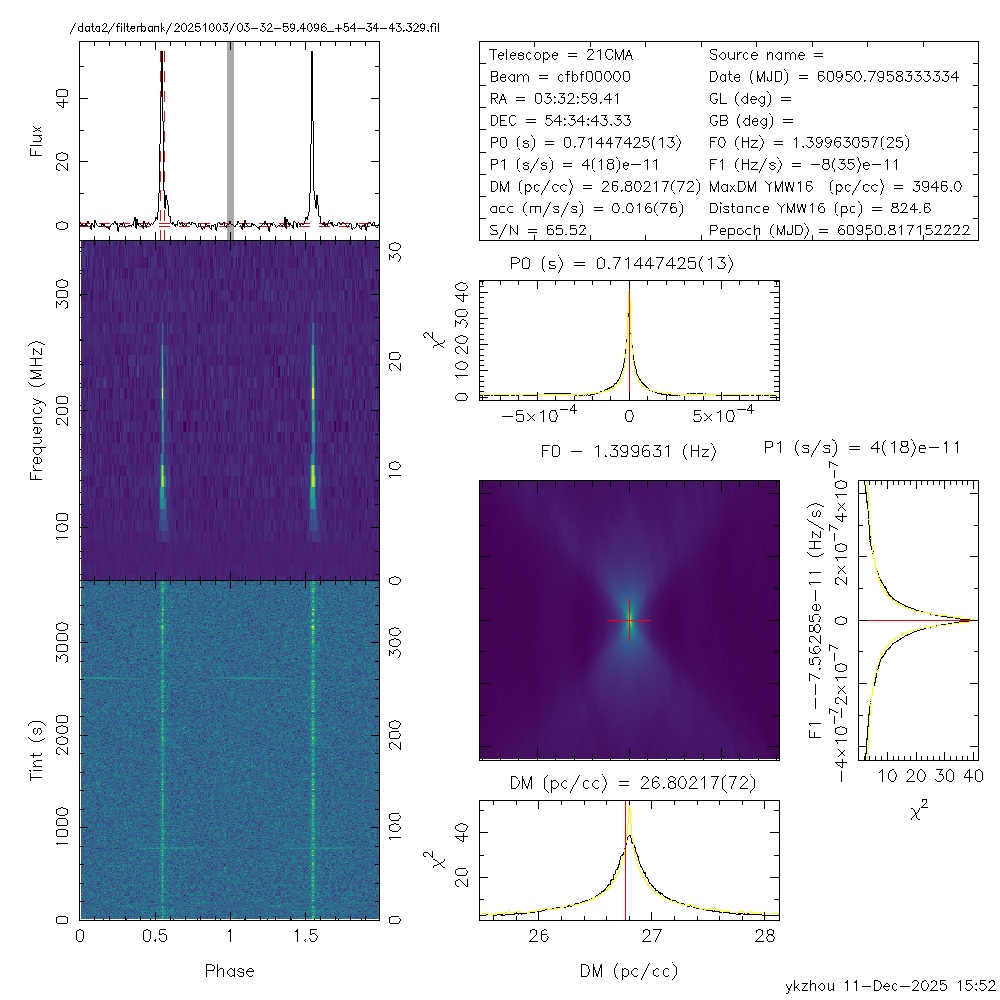}
  \caption{{
    PSR B0329$+$54 observed by a single 21CMA station S13, processed by PulsarX
  }}
  \label{fig:B0329+54_20251003_S13}
\end{figure}

After comparison, MicroPhase ANTSDR T510 is chosen 
as the currently used model for 21CMA pulsar backend,
with deployment shown in \autoref{fig:deployed_rfsoc}.

A test observation was carried out targeting PSR B0329$+$54
using a single 21CMA station S13.
It started at 2025-10-02 19:06:00 UTC and lasted for about 1 hour.
The sample rate was 1600 Msps with decimation by a factor of 2,
800 Msps effectively.
All eight data streams were enabled but only one stream was recorded;
no packet loss was observed.

Baseband data was then processed by a custom program
to form a SIGPROC\footnote{\url{https://sigproc.sourceforge.net/sigproc.pdf}} filterbank file.
The frequency range of this file is trimmed from 0 -- 400 MHz to 50 -- 350 MHz
to avoid memory overflow caused by
the dedispersion algorithm on frequency channels near 0 MHz
and 
to remove frequency components outside the effective bandpass of the system.
This filterbank file is then folded by PulsarX \citep{PulsarX}\footnote{\url{https://github.com/ypmen/PulsarX}}
with parameters provided by the ATNF Pulsar Catalog \citep{psrcat}\footnote{\url{https://www.atnf.csiro.au/research/pulsar/psrcat/}}.
The diagnostic figure, \autoref{fig:B0329+54_20251003_S13}, shows that
PSR B0329$+$54 is detected.

\subsection{Multi-board Synchronization Verification and Initial Calibration of Partial 21CMA Stations}
\label{section:verification_of_synchronization_and_initial_calibration}

As described in \autoref{section:basic_information_of_21CMA},
signals of 21CMA stations are transmitted via radio-over-fiber transceivers
from each station to the central processing room, but the lengths of the optical fibers are not equal;
calibration using bright radio sources is thus required to determine these unknown lengths.

Two persistent radio sources in the northern sky, Cassiopeia A and Cygnus A,
are used as the main calibration sources to obtain signal delays between stations.
Taurus A (Crab Nebula) was also tested, but no obvious signal coherence among station pairs was shown,
possibly because of the antenna response
since its declination (approximately $ +22^{\circ} $) is too far away from
the phase center of a single log-periodic antenna (declination $ + 90^{\circ} $),
hence it was dropped from the calibration source list;
for the same reason, some other sources such as the Sun and Jupiter are also not used for now.

\begin{figure}[!htbp]
  \centering
  \includegraphics[width=1\textwidth]{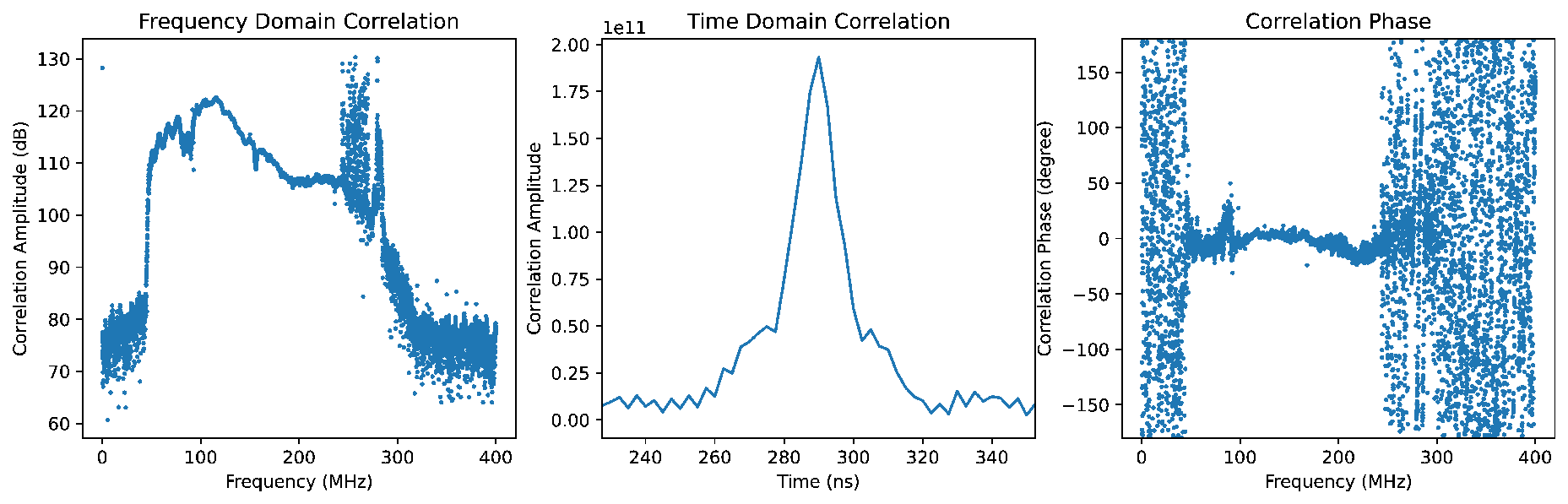}
  \caption{{
      Example of a delay fitting for S07-S03 baseline using Cas A at 2025-12-03 13:00:00 UTC.
      Left: X-axis is sky frequency in MHz, Y-axis is uncalibrated amplitude.
      Middle: X-axis is relative time in nanoseconds, zoomed in around the peak position of uncalibrated amplitude. Y-axis is uncalibrated amplitude.
      Right: X-axis is frequency in MHz, Y-axis is the phase of frequency-domain correlation in degrees,
      with phase difference caused by signal delays fitted and removed.
  }}
  \label{fig:find_delay_example}
\end{figure}

An example of signal delay fitting is shown in \autoref{fig:find_delay_example}.
Procedures are as follows:
for time-domain baseband sample values $ x(t) $ from two stations $ i $, $ j $,
frequency-domain correlation $ C(f) $ is calculated as
\begin{equation}
  C_{i, j}(f) = \big(\mathcal{F}\{x_i(t)\}\big)^{*} \mathcal{F}\{x_j(t)\}
\end{equation}
and time-domain correlation
\begin{equation}
  C_{i, j}(t) = \mathcal{F}^{-1}\{C_{i, j}(f)\}
\end{equation}
where $ \mathcal{F}\{\cdot\} $ is Fourier transformation,
$ (\cdot)^{*} $ is complex conjugate;
these correlations are shown in the left and middle sub-figures of \autoref{fig:find_delay_example}.
The position of time-domain correlation with maximum amplitude is found
and denoted as $ t_{i, j, C(t), \text{max}} $;
it indicates the integer part of the signal delay difference in time domain between station $ i $ and $ j $.
Initial correction of frequency-domain correlation is done as
\begin{equation}
  C'_{i, j}(f) = C_{i, j}(f) \times \mathrm{e}^{2\pi \mathrm{i} (f / f_{\text{max}}) t_{i, j, C(t), \text{max}}}
\end{equation}
where $ f_{\text{max}} $ is the maximum of observation frequency, here 400 MHz.
For fractional signal delay,
Random Sample Consensus (RANSAC) algorithm \citep{RANSAC} is applied to $ C'_{i, j}(f) $
to obtain a robust linear fit;
let the slope be denoted as $ k_{i, j} $,
and then the difference of time-domain signal delay is
\begin{equation}
  \Delta t_{i, j} = t_{i, j, C(t),\text{max}} - \frac{k_{i, j}}{2 \pi}
\end{equation}
the corresponding effective optical path length is
\begin{equation}
  \Delta L_{i, j} = c \Delta t_{i, j}
\end{equation}
where $ c $ is the speed of light.
After removing the contribution of optical path lengths given above,
the phase of the frequency-domain correlation is shown in right sub-figure of \autoref{fig:find_delay_example}.

  When considering multiple baselines,
  signal delays are simply modeled as geometric delay and optical fiber (``cable'') delay.
  For station $ i $, $ j $,
  let the position be denoted $ \vec{r} $ and
  the direction of the source $ \vec{d} $,
  \begin{equation}
    \Delta t_{i, j} = (\vec{r}_i - \vec{r}_j) \cdot (- \vec{d}) + (\Delta t_{\text{cable}, i} - \Delta t_{\text{cable}, j})
  \end{equation}
  where the former term is the difference of geometric delay, and
  the latter term is the difference of cable delay.

  In practice, for the eight stations currently available,
  two MicroPhase ANTSDR T510 sample boards are used for data sampling,
  and the synchronization signals are connected as in \autoref{fig:rfsoc_system_design}.
  A script is written to fit cable delay $ \Delta t_{\text{cable}, i} $
  using the least squares method with the loss function set to ``Cauchy'' to overcome outliers;
  most outliers are caused by severe radio frequency interference.
  For stations currently available, S07 is chosen as the reference due to its geometric position.

\begin{figure}[!htbp]
  \centering
  \includegraphics[width=1\textwidth]{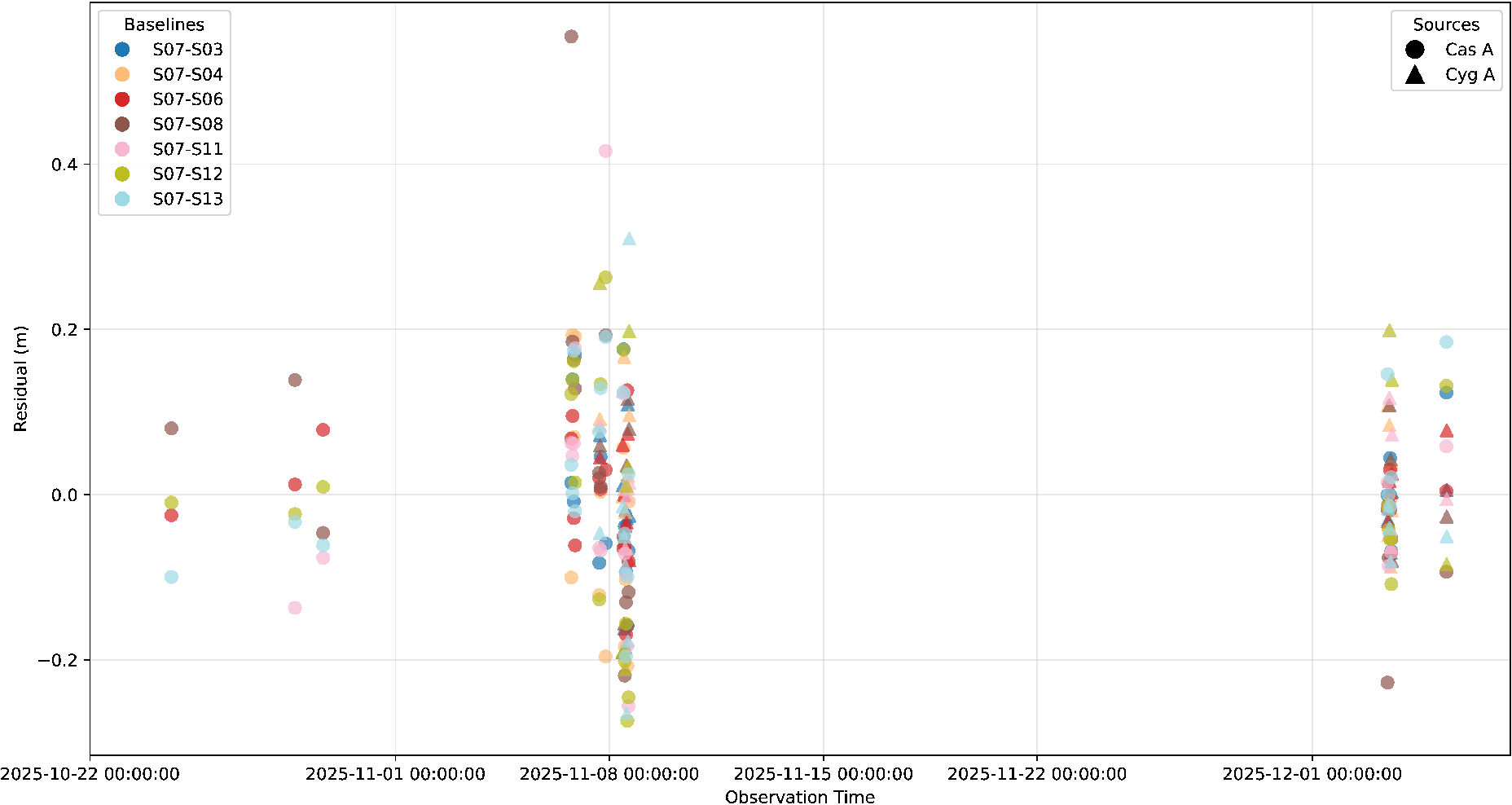}
  \caption{{
    Residuals of the fitted observed signal delay
    using a model consisting only of geometric delay and cable delay.
    X-axis: date and time of observations.
    Y-axis: residual of signal delay, i.e. difference of actual delay and modelled delay.
    Colors represent baselines, shapes represent target sources.
  }}
  \label{fig:fit_delay_20251024_20251205}
\end{figure}

  By monitoring signal delays over a timescale of months,
  cable delay solutions relative to S07 of stations S03, S04, S06, S08, S11, S12, S13 were obtained, shown in \autoref{table:cable_delay}.
  Other stations were excluded because they are still under construction at the time of writing.
  Successful calibration observations and their residuals are shown in \autoref{fig:fit_delay_20251024_20251205}.

  \begin{table}
    \centering
    \caption{{
      Station cable delay solutions from observations shown in \autoref{fig:fit_delay_20251024_20251205}.
      Note that uncertainties given in this table only account for statistical uncertainties and do not contain systematic uncertainties.
    }}
    \begin{tabular}{ccccc}
      \hline
      Baseline  & S07 -- S03               & S07 -- S04                & S07 -- S06               & S07 -- S08               \\
      Delay (m) & $ 265.837 \pm 0.017 $ & $ -336.575 \pm 0.023 $ & $ -96.632 \pm 0.011 $ & $ 210.881 \pm 0.028 $ \\
      \hline
      Baseline  & S07 -- S11               & S07 -- S12                & S07 -- S13               &                       \\
      Delay (m) & $ 355.045 \pm 0.022 $ & $ 405.509 \pm 0.028 $  & $ 451.507 \pm 0.022 $ &                       \\
      \hline
    \end{tabular}
    \label{table:cable_delay}
  \end{table}

  Delay values in \autoref{table:cable_delay} provide essential parameters
  for beamforming.
  \autoref{fig:fit_delay_20251024_20251205} shows that
  the residual of this simple model is mostly within one quarter wavelength of the highest frequency considered,
  hence these delays are considered accurate enough for use.

\begin{figure}[!htbp]
  \centering
  \includegraphics[width=1\textwidth]{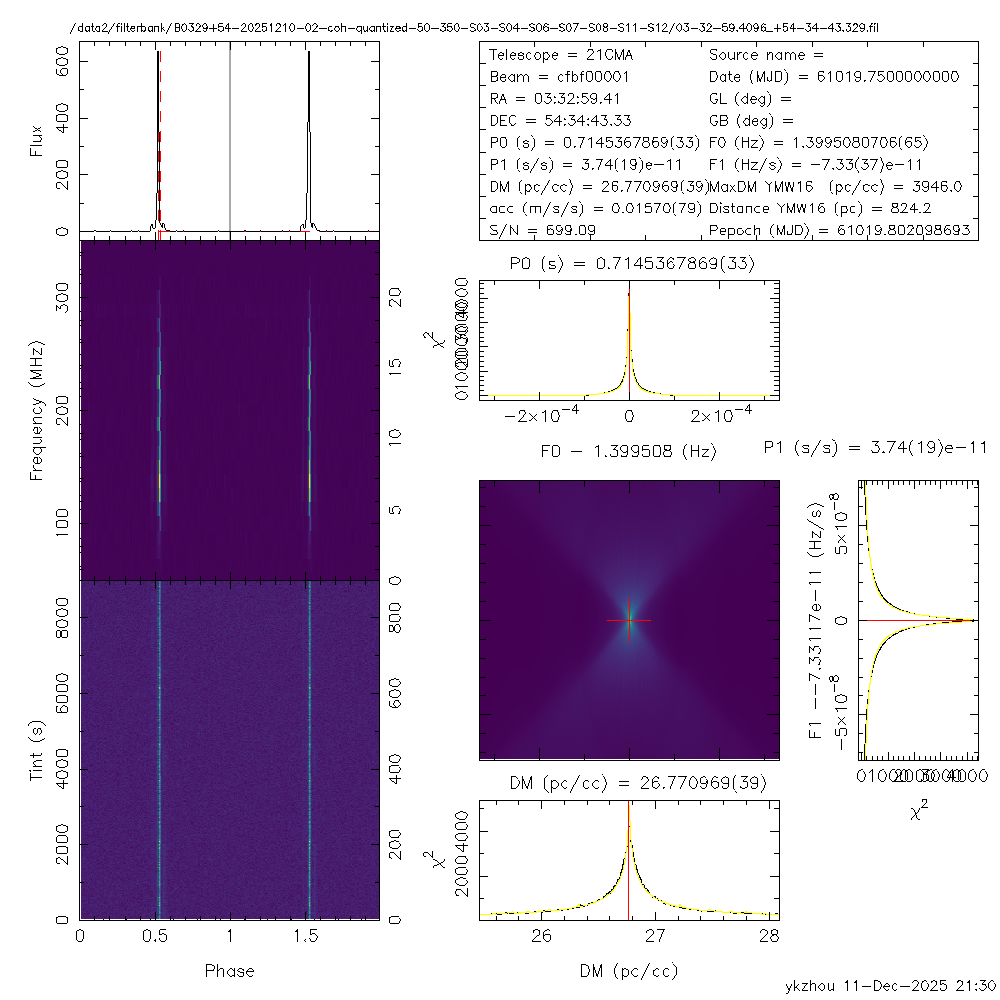}
  \caption{{
    Observation of PSR B0329$+$54 
    with delay values given in \autoref{table:cable_delay},
    processed by PulsarX
  }}
  \label{fig:B0329+54-20251210-PulsarX}
\end{figure}

\begin{figure}[!htbp]
  \centering
  \includegraphics[width=0.5\textwidth]{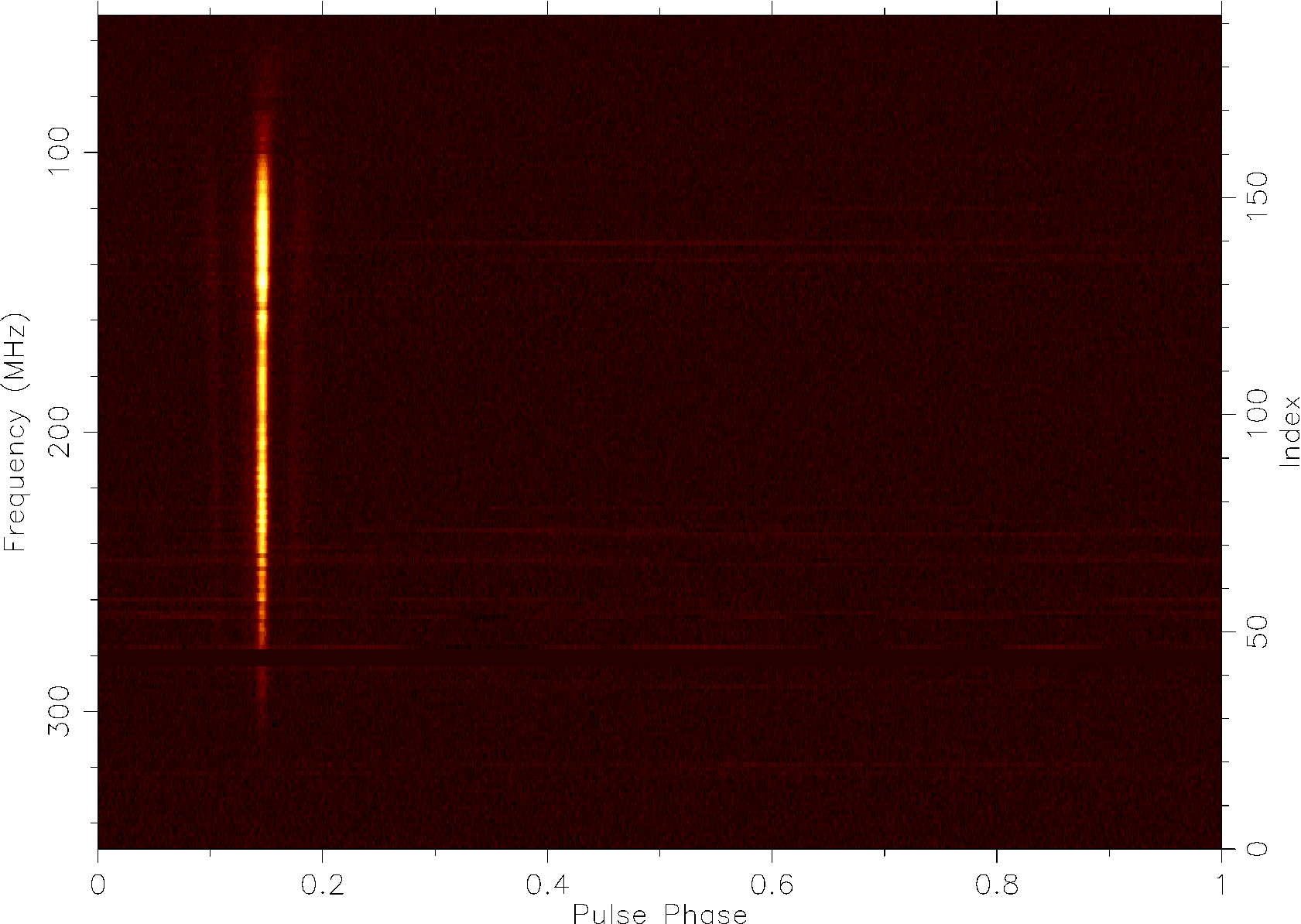}
  \caption{{
    Observation of PSR B0329$+$54 
    with delay values given in \autoref{table:cable_delay},
    processed by DSPSR and PSRCHIVE.
    X-axis is the spin phase of this pulsar, Y-axis is observation frequency in MHz,
    color represents relative intensity of the folded signal.
  }}
  \label{fig:B0329+54-20251210}
\end{figure}

  An example of coherent beamforming using delay values given in \autoref{table:cable_delay}
  is shown in \autoref{fig:B0329+54-20251210-PulsarX} and \autoref{fig:B0329+54-20251210}.
  This observation started at 2025-12-10 18:00:00 UTC
  and lasted for 2.5 hours.
  The stations used are S03, S04, S06, S07, S08, S11, S12.
  Signals from these stations were processed by a custom beamforming program
  to form a SIGPROC filterbank file.
  This filterbank file was then folded by PulsarX and
  DSPSR \citep{dspsr} \footnote{\url{https://dspsr.sourceforge.net/}},
  and then processed by 
  PSRCHIVE \citep{psrchive} \footnote{\url{https://psrchive.sourceforge.net/}}.
  The signal from this pulsar is detected in 65 -- 305 MHz.
  This joint observation with Yunnan Astronomical Observatory
  investigates multi-frequency mode switching of PSR B0329$+$54
  and will be discussed in another paper (Shen et al. in prep.).
  Detailed explanation of this GPU- and RDMA-based multi-node data processing pipeline,
  including channelization stage (F-engine), correlator (X-engine), and beamformer (B-engine)
  will be presented in an upcoming paper.

\section{Discussion}
\label{sect:discussion}

\subsection{Possible Origins of Inter-Station Delay Calibration Errors in 21CMA}

In the laboratory measurements (\autoref{section:lab_test}), 
we demonstrated that the calibration accuracy of inter-board delays is better than the sampling period. 
Nevertheless, as shown in \autoref{section:verification_of_synchronization_and_initial_calibration}
(\autoref{fig:find_delay_example} and \autoref{fig:fit_delay_20251024_20251205}), 
corrections of inter-station delays fail to fully remove the phase discrepancies across frequency channels. 
The residual calibration errors may be caused by a variety of other factors.

The first possible source may lie within the analog signal chain 
described in \autoref{section:basic_information_of_21CMA};
this chain may introduce non-flat group delays.
The delay calibration described above considers only optical path length and
adopts a linear phase-frequency model, which effectively ignores any non-flat group delays;
this may explain the residuals in the inter-station delay fitting shown in \autoref{fig:find_delay_example}. 

Another potential source of error arises from 
the use of analog delay units for delaying and summing antenna signals. 
When the beam tracks a celestial target, 
the effective gains of individual antennas within a station may still vary despite prior compensation, 
leading to shifts in the effective coordinates of the stations,
an effect that is not accounted for in the calibration process. 
This may explain the drifts in inter-station delay observed in \autoref{fig:fit_delay_20251024_20251205}.

\subsection{Design Specifications of Acquisition Units for Radio Telescopes}
Development of the 21CMA pulsar data acquisition system shows some requirements on hardware design of sample boards
for low- and mid-frequency radio astronomy,
as follows.

\begin{enumerate}
    \item The frequency of the voltage-controlled oscillator (VCXO) for the clock chip
          should be carefully selected to allow greater flexibility in generating the sampling clock
          when working in conjunction with the external 10 MHz global reference clock.
          For example, 122.88 MHz crystal oscillators are commonly employed for 4G and 5G cellular communication, 
          but compared to 100 MHz oscillators, 122.88 MHz ones are suboptimal in radio astronomy as they significantly reduce the frequency of the phase detector in a phase-locked loop.
    \item Digital signal inputs with voltage standards uncommon for FPGAs, especially the 5 V TTL 1PPS signal, 
          should be fed to the board through proper buffers to avoid damage to the core chip.
    \item Two QSFP interfaces should be provided to fully utilize the sampling and transmission capabilities of the RFSoC
          and to improve applicability to scenarios with high data rates, 
          e.g., large-scale radio interferometers, ultra-wideband receivers with many intermediate-frequency bands,
          and massive multiple-input multiple-output (MIMO) in phased array feeds (PAFs).
    \item All ADCs should be brought out to radio frequency connectors,
          to improve applicability to scenarios same with the previous point.
\end{enumerate}

\section{Conclusions}
\label{sect:conclusion}

  In this work, we implemented and deployed an RFSoC-based data acquisition system for 21CMA.
  Developed using a block design, it samples baseband signals and
  transmits them directly over 100 Gb Ethernet,
  with instantaneous bandwidth up to 400 MHz, of which 50 -- 350 MHz is effective for 21CMA.
  The packet format is compatible with FAST's ROACH2-based multibeam backend.
  Multi-board synchronization and stability of the whole system
  are verified over a timescale of several months.
  Initial results of single-station observation of pulsar B0329$+$54,
  signal delay calibration, and multi-station observation using
  coherent beamforming are presented,
  demonstrating a possibility of low-frequency wideband census studies and large-scale pulsar surveys.

\begin{acknowledgements}
  This work is supported by the
  National SKA Program of China (2020SKA0120200, 2020SKA0120100, 2020SKA0110200)
  and the National Science Foundation of China (Grant No. 12573097).

  The authors sincerely appreciate Prof. Dan Werthimer and the CASPER group for their open-source hardware platforms and firmware designs.
  We would like to thank Yan Zhu and Rushuang Zhao for support during development of this system,
  Xun Shi for discussion,
  Gregory Herczeg and Zu Yan for editing this paper in the course ``Advanced Writing for Astronomy''.
\end{acknowledgements}

\appendix                  

\bibliographystyle{raa}
\bibliography{bibli}

\label{lastpage}

\end{document}